\documentclass[letter,twocolumn,a4paper]{jpsj3}
\usepackage{bm}
\usepackage{graphicx}
\usepackage{subfigure}

\title{Three-Dimensional Dirac Electrons at the Fermi Energy in Cubic
Inverse Perovskites: Ca$_3$PbO and Its Family}
\author{Toshikaze \textsc{Kariyado}\thanks{E-mail address: kariyado@hosi.phys.s.u-tokyo.ac.jp} 
and Masao \textsc{Ogata}}
\inst{Department of Physics, the University of Tokyo, 7-3-1 Hongo, Bunkyo, Tokyo
113-0033, Japan}

\abst{
The band structure of cubic inverse perovskites, Ca$_3$PbO and its
family, are investigated with the first-principles method. A close
observation of the band structure reveals that six
equivalent Dirac electrons with a very small mass exist on the line
connecting the $\Gamma$- and X-points, and at the symmetrically
equivalent points
in the Brillouin zone. The discovered Dirac electrons are
three-dimensional and remarkably
located exactly at the Fermi energy. A tight-binding model
describing the low-energy band structure is also
constructed and used to discuss the origin of the
Dirac electrons in this material. 
Materials related to Ca$_3$PbO are also studied, and some
design principles for the Dirac electrons in this series of materials
are proposed. 
}

\kword{Dirac electron, inverse perovskite, the first-principles
calculation, tight-binding model}

\begin{document}
\maketitle

``Emergence'' is one of the most important concepts in 
condensed matter physics. Although this key word often
appears in the context of many-body or strong correlation effects,
emergent
behaviors are also observed in noninteracting systems where the
low-energy effective Hamiltonian becomes quite distinct from the
original Hamiltonian. The
most well-known example is the relativistic Dirac Hamiltonian realized in
graphene \cite{Novoselov:2005fk} derived from a nonrelativistic
Hamiltonian. Many intriguing properties of
graphene can be ascribed to the existence of ``Dirac
electrons'' in its low-energy band structure\cite{Geim:2007uq}. 
Actually, Dirac electrons in materials have a long history starting from 
bismuth, which has three-dimensional massive Dirac electrons in its
band structure\cite{Wolff19641057}. 
The organic conductor $\alpha$-(BETT-TTF)$_2$I$_3$ is also known to be a
material having Dirac electrons near the Fermi
energy\cite{JPSJ.75.054705,alpha_review}. The most up-to-date example is a
surface state of a three-dimensional
topological insulator\cite{RevModPhys.82.3045}, which is extensively
studied in these days.

In connection with topological insulators, inverse-perovskite
materials have attracted much attention recently. For example, it is
claimed that Ca$_3$NBi enters a topological phase under an
appropriate strain engineering scheme\cite{PhysRevLett.105.216406}. 
In this paper, we show that cubic inverse perovskites, Ca$_3$PbO and its
family, have three-dimensional Dirac electrons with a very small mass at
the Fermi energy. Although Ca$_3$PbO is on
the list of potential topological insulators proposed by
Klintenberg\cite{Klintenberg:2010uq}, our close
observation of its band structure reveals the existence of bulk (not
surface) Dirac electrons on the line connecting the $\Gamma$- and
X-points, and at the symmetrically equivalent points in the Brillouin
zone. Although some first-principles calculations on this material are
available in the literature\cite{Haddadi20101995,Ortiz20091042}, it
is the first time that the existence of Dirac electrons is pointed
out. We also construct a tight-binding model that
describes the low-energy band structure, and clarify the
origin of the Dirac electrons in this material by
analyzing the model. We also study the family
of Ca$_3$PbO, and some design principles for Dirac electrons in this
series of materials are proposed on the basis of the obtained results. 

\begin{figure}[tbp]
 \begin{center}
  \includegraphics[scale=1.0]{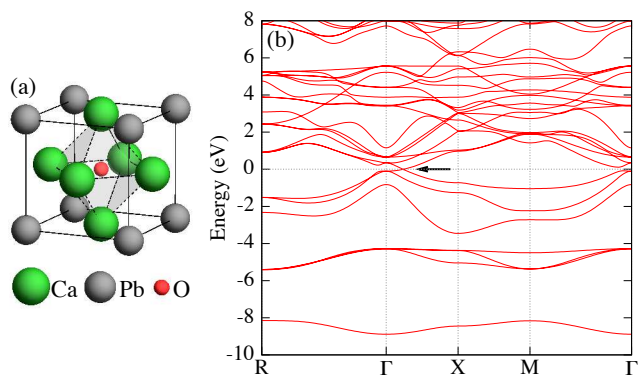}
  \caption{(Color online) (a) Cubic inverse-perovskite structure of
  Ca$_3$PbO. (b) The calculated band structure of Ca$_3$PbO
  along the high symmetry lines in the Brillouin zone. The location of
  one of the Dirac points is marked with an arrow.}
  \label{fig1}
 \end{center}
\end{figure}

The crystal structure of Ca$_3$PbO is shown in Fig.~\ref{fig1}(a). It
belongs to the space group $Pm\bar{3}m$\cite{Widera19801805} and is an
inverse perovskite that possesses an O atom surrounded octahedrally
by Ca atoms. 
Figure~\ref{fig1}(b) shows the band structure of Ca$_3$PbO obtained
within the first-principles calculation using the WIEN2k
package\cite{wien2k}, in which
the full-potential augmented-plane-wave method is implemented. 
The spin-orbit coupling is
taken account of within the muffin-tin sphere of each atom via the second
variational step\cite{0022-3719-13-14-009}. 
The required crystal parameters are taken from experimental
results\cite{Widera19801805}. 
$20\times 20\times 20$
$k$-points, which result in 220 $k$-points in the reduced Brillouin zone,
are employed in the self-consistent cycle of our calculations. The charge
density obtained with this number of $k$-points is used to calculate band
structures on the finer momentum meshes required to confirm the existence
of Dirac electrons.
An overview of the calculated band structure shown in
Fig.~\ref{fig1}(b) is as follows. First, the band at approximately $-8$
eV and the bands
between $-4$ and $0$ eV mainly originate from
Pb $6s$ and Pb $6p$ orbitals, respectively. 
On the other hand, the bands between $-6$ and $-4$ eV originate from O $2p$
orbitals, and the highly entangled bands between $0$ and $6$ eV mostly
originate from Ca $3d$ orbitals. These observations mean that Pb $6s$
and O $2p$ orbitals are completely filled, and Pb $6p$ orbitals are almost
completely filled, which is roughly consistent with
Ca$^{2+}_3$Pb$^{4-}$O$^{2-}$. 

\begin{figure}[tbp]
 \begin{center}
  \includegraphics[scale=1.0]{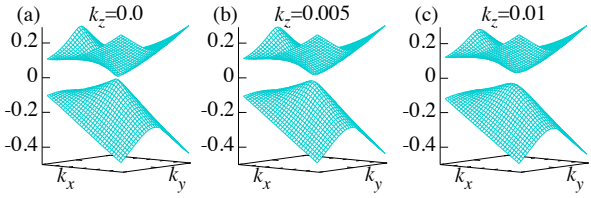}
  \caption{(Color online) Dispersion relations around one of the Dirac
  points for 
  $(k_x,k_y,k_z)$ where $0.04\leq k_x\leq 0.2$, $-0.08\leq k_y \leq 0.08$,
  and $k_z$ has several values. The wave numbers are in the unit of
  $2\pi/a_0$, with $a_0$ being the lattice constant.}
  \label{fig2}
 \end{center}
\end{figure}
A Dirac point is a $\bm{k}$-point in the Brillouin zone around which
Dirac electrons exist. We find that a Dirac point exists on the
high-symmetry line connecting the $\Gamma$- and X-points, and its
location is marked with an arrow in
Fig.~\ref{fig1}(b). Note that the marked point is exactly at the Fermi
energy. Owing to the cubic symmetry, six Dirac
points exist at the symmetrically equivalent points in the Brillouin
zone.
In order to clarify the dispersion relations, three-dimensional plots of
the band structure near the Dirac point are shown in Fig.~\ref{fig2}.
We can see a conelike dispersion relation on
the $k_x$-$k_y$ plane when $k_z=0$ [Fig.~\ref{fig2}(a)], and the gap between
the valence band and the conduction
band increases away from $k_z=0$ [Figs.~\ref{fig2}(b) and
\ref{fig2}(c)]. These findings confirm the existence of
three-dimensional Dirac electrons in the band structure. Actually, even for
$k_z=0$ [Fig.~\ref{fig2}(a)], the
first-principles calculation gives a small gap of about $14$ meV. Namely,
the discovered Dirac electrons are {\it massive}. However, the estimated
mass gap of about $14$ meV is sufficiently small to observe physical phenomena
associated with the
linear dispersion of a Dirac electron. In fact, Fig.~\ref{fig2}(a) shows
that the linear (or sublinear) dispersion extends up to an energy higher
than $0.2$ eV, which is one order greater than the gap.

\begin{figure}[tbp]
 \begin{center}
  \includegraphics[scale=1.0]{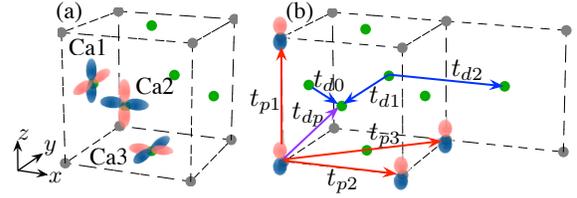}
  \caption{(Color online) (a) Naming convention of the three Ca sites and
  schematic view of the $d$-orbitals included in the basis set. (b)
  Definitions of hopping integrals. In both (a) and (b), the Pb sites are at
  the corners of the cube and the Ca sites are at the center of the
  surfaces.}
  \label{fig3}
 \end{center}
\end{figure}

In order to understand the emergence of the Dirac electrons in this
material, we construct a tight-binding model describing the low-energy
band structure. By analyzing the
orbital-weight distribution in the obtained band
dispersions\cite{comment1} in Fig.~\ref{fig1}(b), we find
that the bands forming the Dirac electrons mainly originate from the Pb
$6p$ orbital and one type of Ca $3d$ orbital. It will become important
later that an overlap exists between these $p$- and $d$-bands, i.e., the top of
the $p$-bands (located at the $\Gamma$-point) has an energy of 
$\sim 0.3$ eV while the bottom of the $d$-bands (located at the
$\Gamma$-point) has an energy of $\sim -0.1$ eV.
Among the Ca $3d$ orbitals, those
relevant to the Dirac electrons are the $d_{y^2-z^2}$ orbital on the Ca1
site, the $d_{z^2-y^2}$ orbital on the Ca2 site, and the $d_{x^2-y^2}$ orbital
on the Ca3 site, all of which are equivalent in the cubic symmetry.
Figure \ref{fig3}(a) shows the definition of the Ca1, Ca2, and Ca3 sites
and the description of the above three orbitals.
Based on these results, we use the 12 local orbitals
\begin{align}
 |p_x\sigma\rangle, && |p_y\sigma\rangle, && |p_z\sigma\rangle, &&
 |d_1\sigma\rangle, && |d_2\sigma\rangle, && |d_3\sigma\rangle, 
 \label{basis_1}
\end{align} 
($\sigma=\uparrow$, $\downarrow$) as a basis set for our tight-binding
model, where $d_1$, $d_2$, and $d_3$
represent the above-mentioned $d$-orbitals on the Ca1, Ca2, and Ca3 sites,
respectively. Assuming the nearest-neighbor and next-nearest-neighbor
hoppings shown in Fig.~\ref{fig3}(b), we have the tight-binding Hamiltonian
\begin{multline}
 \hat{H}
  =
  \sum_\sigma\sum_{\bm{r}p}\epsilon_p c^\dagger_{\bm{r}p\sigma}c_{\bm{r}p\sigma}
  +\sum_\sigma\sum_{\bm{r}d}\epsilon_d c^\dagger_{\bm{r}d\sigma}c_{\bm{r}d\sigma}\\
  +\sum_\sigma\sum_{\bm{r}a\bm{r}'a'}
  t^{\sigma\sigma'}_{aa'}(\bm{r}-\bm{r}')
  c^\dagger_{\bm{r}a\sigma}c_{\bm{r}'a'\sigma}
  +2\lambda\sum_{\bm{r}}\bm{l}_{\bm{r}}\cdot\bm{s}_{\bm{r}},
  \label{Hamiltonian0}
\end{multline}
where indices $a$ and $a'$ represent $p_{x,y,z}$ or $d_{1,2,3}$, and the
fourth term represents the spin-orbit coupling for Pb atoms. Owing to the
Pb $p$-orbital character, $|p_x\uparrow\rangle$,
$|p_y\uparrow\rangle$, and
$|p_z\downarrow\rangle$ are mixed. Here the spin-orbit couplings for Ca
atoms are neglected for simplicity. The Fourier
transformation of
eq.~\eqref{Hamiltonian0} gives
\begin{equation}
  \hat{H}
  =\sum_{\bm{k}}\sum_{\alpha\alpha'}\mathcal{E}_{\alpha\alpha'}(\bm{k})
  c^\dagger_{\bm{k}\alpha}c_{\bm{k}\alpha'}.
  \label{Hamiltonian1}
\end{equation}
After carrying out the Fourier transformation, we transform the matrix
elements by attaching the momentum-dependent phase factors to the basis
orbitals as 
$|p_{x,y,z}\sigma\rangle\rightarrow \mathrm{e}^{i(k_x+k_y+k_z)/2}|p_{x,y,z}\sigma\rangle$, 
$|d_1\sigma\rangle\rightarrow \mathrm{e}^{ik_x/2}|d_1\sigma\rangle$,
$|d_2\sigma\rangle\rightarrow \mathrm{e}^{ik_y/2}|d_2\sigma\rangle$, and
$|d_3\sigma\rangle\rightarrow
\mathrm{e}^{ik_z/2}|d_3\sigma\rangle$. This transformation makes the
expressions for the matrix elements simple, and the transformed basis and
matrix elements are used in the following.

\begin{figure}[tbp]
 \begin{center}
  \includegraphics[scale=1.0]{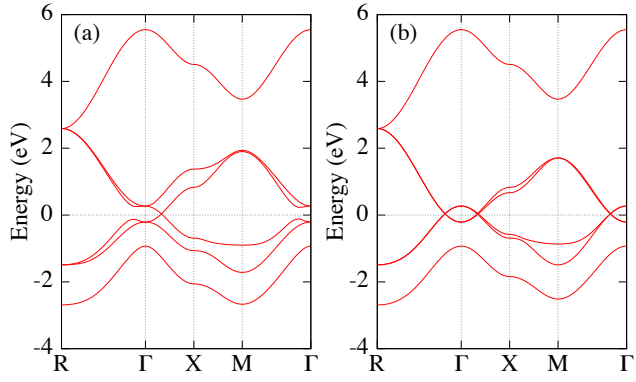}
 \caption{(Color online) Band structures obtained
  from the tight-binding model [eq.~\eqref{Hamiltonian0}] with (a) and
  without (b) the hybridization between $p$-orbitals and $d$-orbitals.}
  \label{fig4}
 \end{center}
\end{figure}

Figure~\ref{fig4}(a) shows the band structure of our
tight-binding model obtained using the parameter set $\epsilon_p=-1.25$,
$t_{p1}=0.20$, $t_{p2}=0.12$, $t_{p3}=0.06$, $\epsilon_d=2.15$,
$t_{d0}=0.48$, $t_{d1}=0.22$, $t_{d2}=-0.22$, $t_{dp}=0.22$, and
$\lambda=0.4$ (in units of eV). 
Although there are some differences between the first-principles bands
and the tight-binding bands, the latter bands reproduce the
important features (including the Dirac electrons) of the former
bands having large Ca1 $d_{y^2-z^2}$, Ca2 $d_{z^2-y^2}$, Ca3 $d_{x^2-y^2}$,
and Pb $p$-orbital weights. In this simplified model, however, there is
no gap at the Dirac point, i.e., we obtain
{\it massless} Dirac electrons. It is necessary
to include Ca $d_{xy}$ or $d_{zx/yz}$ orbitals to have a finite mass
term. (Details will be published elsewhere.)

In order to obtain a clearer view on the emergence of the Dirac electrons,
we derive a low-energy effective Hamiltonian of the above
tight-binding model and show that it really results in a Dirac Hamiltonian.
Firstly, by introducing a new basis set, we eliminate the uppermost and
lowermost bands, which are well separated from the bands forming Dirac
electrons. Since the Dirac point is close to
the $\Gamma$-point, we use the new basis set, which consists of the
eigenstates at the $\Gamma$-point. At the $\Gamma$-point, we can treat
the $p$- and
$d$-orbitals separately since the hybridization between them becomes
zero. For the $p$-orbitals, the eigenvalues are
$\epsilon_{p0}+\lambda$ and $\epsilon_{p0}-2\lambda$, where
$\epsilon_{p0}\equiv \epsilon_p+2t_{p1}+4t_{p2}+4t_{p3}$. We find that
the former corresponds to the states forming the Dirac electrons and the
latter to
the lowermost band. The energy splitting between them is
caused by the spin-orbit coupling, and the former has a total angular
momentum of $j=3/2$. The new basis set for $j=3/2$ states is
\begin{subequations}
 \begin{align}
 |p_{\frac{3}{2}\frac{3}{2}}\rangle&=
 -\bigl(|p_x\uparrow\rangle+i|p_y\uparrow\rangle\bigr)/\sqrt{2},\\
 |p_{\frac{3}{2}\frac{1}{2}}\rangle&=
 -\bigl(|p_x\downarrow\rangle+i|p_y\downarrow\rangle-2|p_z\uparrow\rangle\bigr)/\sqrt{6},\\
  |p_{\frac{3}{2}\bar{\frac{1}{2}}}\rangle&=
  \bigl(|p_x\uparrow\rangle-i|p_y\uparrow\rangle+2|p_z\downarrow\rangle\bigr)/\sqrt{6},\\
  |p_{\frac{3}{2}\bar{\frac{3}{2}}}\rangle&=
  \bigl(|p_x\downarrow\rangle-i|p_y\downarrow\rangle\bigr)/\sqrt{2}.
 \end{align}\label{new_basis_p}\end{subequations}
For the $d$-orbitals, the eigenvalues are
$\epsilon_{d0}-4t_{d0}$ and $\epsilon_{d0}+8t_{d0}$, where
$\epsilon_{d0}\equiv \epsilon_d+2t_{d1}+4t_{d2}$. The former corresponds
to the bands near the Fermi energy and latter to the uppermost
band. Explicitly, the wave functions for the former are
\begin{subequations}
 \begin{align}
  |d_1'\sigma\rangle&=
  \bigl(|d_1\sigma\rangle-|d_2\sigma\rangle\bigr)/\sqrt{2},\\
  |d_2'\sigma\rangle&=
  \bigl(|d_1\sigma\rangle+|d_2\sigma\rangle-2|d_3\sigma\rangle\bigr)/\sqrt{6}.
 \end{align}\label{new_basis_d}\end{subequations}
Performing a unitary transformation from the basis set of eq.~\eqref{basis_1}
into that containing eqs.~\eqref{new_basis_p} and \eqref{new_basis_d},
we transform the $12\times 12$ matrix, $\tilde{\mathcal{E}}_{\bm{k}}$, in
eq.~\eqref{Hamiltonian1} into $\tilde{\mathcal{E}}_{\bm{k}}'$. Then, we
keep only the matrix elements between the above eight states in
eqs.~\eqref{new_basis_p} and \eqref{new_basis_d}, reducing the matrix
$\tilde{\mathcal{E}}_{\bm{k}}'$ into an $8\times 8$ matrix. Finally, 
concentrating on the Dirac point on the $k_z$-axis, we expand the matrix
elements with respect to $k_x$ and $k_y$ up to the first order in $k_x$ and
$k_y$. After some algebra, we obtain
\begin{equation}
 \hat{\mathcal{E}}''_{\bm{k}}
  =
  \begin{pmatrix}
   g^{pp}_1&0&0&0&-c_1k_-&c_2k_+&0&0\\
   &g^{pp}_2&0&0&c_3&0& -c_1'k_-&c_2'k_+\\
   & &g^{pp}_2 &0 &c_1'k_+ &-c_2'k_- &c_3 & 0\\
   && &g^{pp}_1 &0 &0 &c_1k_+&-c_2k_- \\
   & & & &g^{dd}_{1} &0 &0 & 0\\
   & & & & &g^{dd}_{2} &0 & 0\\
   & & & & & &g^{dd}_{1} &0 \\
   & & & & & & &g^{dd}_{2}
  \end{pmatrix},
  \label{8x8_linear}
\end{equation}
where $k_{\pm}=k_x+ik_y$, $c_1'=c_1/\sqrt{3}$, $c_2'=c_2/\sqrt{3}$, and
\begin{equation}
\begin{split}
 c_1&=it_{dp}\cos(k_z/2),\quad
 c_2=it_{dp}\bigl(2+\cos(k_z/2)\bigr)/\sqrt{3},\\
 c_3&=-8it_{dp}\sin(k_z/2)/\sqrt{3}.
\end{split}
\end{equation}
Note that now the new basis set is
\{$|p_{\frac{3}{2}\frac{3}{2}}\rangle$,
$|p_{\frac{3}{2}\frac{1}{2}}\rangle$,
$|p_{\frac{3}{2}\bar{\frac{1}{2}}}\rangle$,
$|p_{\frac{3}{2}\bar{\frac{3}{2}}}\rangle$,
 $|d_1'\uparrow\rangle$, $|d_2'\uparrow\rangle$, 
 $|d_1'\downarrow\rangle$, $|d_2'\downarrow\rangle$\}. Note also that
 $g^{pp}_i$ and $g^{dd}_i$ depend on $k_z$.

In the limit of $k_x=k_y=0$, all of the off-diagonal elements in
eq.~\eqref{8x8_linear} vanish except for the elements represented as
$c_3$. The finite $c_3$ induces the
strong band repulsion between bands from
$|p_{\frac{3}{2}\frac{1}{2}}\rangle$ and
$|d_1'\uparrow\rangle$ (second and fifth rows), and also between bands from
$|p_{\frac{3}{2}\bar{\frac{1}{2}}}\rangle$ and $|d_1'\downarrow\rangle$
(third and seventh rows). As a result, the bands originating 
from these states are pushed away from the Fermi energy, and thus we
can neglect these states, keeping
only $|p_{\frac{3}{2}\frac{3}{2}}\rangle$,
$|p_{\frac{3}{2}\bar{\frac{3}{2}}}\rangle$,
$|d_2'\uparrow\rangle$, and $|d_2'\downarrow\rangle$ in the following.
Then, suppose that
$g^{pp}_1=g^{dd}_{2}$ is satisfied at some $k_{z0}$. When
$g^{pp}_1=g^{dd}_{2}$ holds, we can expand $g^{pp}_1$ and $g^{dd}_{2}$
as $g^{pp}_1=-c_p\delta k_z+\epsilon_0$ and 
$g^{dd}_2=c_d\delta k_z+\epsilon_0$, where $\delta k_z=k_z-k_{z0}$ and
$\epsilon_0=g^{pp}_1$ ($=g^{dd}_2$) at $k_z=k_{z0}$. 
Using these relations, the Hamiltonian can be written as
\begin{equation}
 \hat{\mathcal{E}}'''_{\bm{k}}=
 (\epsilon_0+\delta c\delta k_z)\hat{1}+
 \begin{pmatrix}
  -c\delta k_z&0&c_2k_+&0\\
  &-c\delta k_z &0 & -c_2k_-\\
  & &c\delta k_z &0\\ 
  & & & c\delta k_z
 \end{pmatrix},
\end{equation}
with $c=(c_d+c_p)/2$ and $\delta c=(c_d-c_p)/2$, 
which gives a tilted massless Dirac Hamiltonian. The Dirac point is
$(0,0,k_{z0})$ in this case.

A very simple and intuitive reason for the emergence of the Dirac
electrons is obtained from the above analysis. Figure~\ref{fig4}(b) shows
the energy dispersion of the
tight-binding model, where $t_{dp}$ is artificially set to zero. This
corresponds to the case that there is no hybridization between the $p$-
and $d$-orbitals. 
In this case, the $p$-bands ($d$-bands) have a usual hole
(electron) Fermi surface around the $\Gamma$-point. Although the $p$-
and $d$-bands overlap with each other, this only leads to a usual band
crossing. If we turn on the hybridization
$t_{dp}$, it generally causes a band repulsion between the $p$- and
$d$-bands. However, owing to the symmetry of the involved orbitals, this
band repulsion does not act on the $\Gamma$-X line. As a result, the band
crossing remains
at an isolated point in the Brillouin zone, leading to the emergence of
a Dirac electron. Note that the existence of an overlap between the $p$-
and $d$-bands supports the assumption $g^{pp}_1=g^{dd}_2$ in the
previous paragraph. Thus, we can state that the overlapping between the
$p$- and $d$-bands is essential for Dirac electrons to appear in this model.

\begin{figure}[tbp]
 \begin{center}
  \includegraphics[scale=1.0]{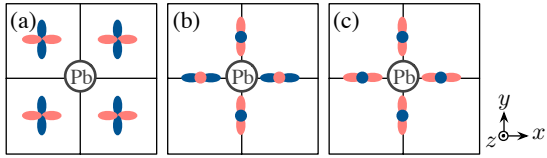}
  \caption{(Color online) Allowed patterns of the
  phase of the wave function for the $d$-orbitals when $k_x=k_y=0$, seen
  from the $z$-axis direction. The dark and light robes represent the
  positive and negative values of the wave function, respectively.
  }
  \label{fig5}
 \end{center}
\end{figure}
Next, we discuss the reason why the band repulsion does not
act on the $\Gamma$-X line. On the $k_z$-axis, which is equivalent to
the $\Gamma$-X line, the $d$-orbital wave function can be classified as
shown in Figs.~\ref{fig5}(a)-\ref{fig5}(c). 
Actually, $|d_1'\sigma\rangle$
corresponds to the wave function in Fig.~\ref{fig5}(c), while
$|d_2'\sigma\rangle$ is
a superposition of those in Figs.~\ref{fig5}(a) and \ref{fig5}(b).
From these figures, we can easily see that all of the hybridization
between the Pb $p$-orbitals and the wave functions in Fig.~\ref{fig5}
vanish except for that between the $p_z$ orbital and the wave function in
Fig.~\ref{fig5}(c). This is
the reason why we have a finite off-diagonal matrix element $c_3$ for
$|d_1'\sigma\rangle$, while the off-diagonal matrix elements
for $|d_2'\sigma\rangle$ vanish as $|k_{\pm}|\rightarrow 0$. The
vanishing matrix elements imply that the band repulsion does not
act. 

Note that, although the above graphical argument is heuristic and
intuitive, the
spin-orbit coupling is not taken into account in this argument. By
analyzing the
irreducible representation of the bands obtained in the first-principles
calculation\cite{comment1}, we find that the two bands forming the
Dirac electrons fall into the same irreducible representation on the
$\Gamma$-X line. As a result, there is a
small hybridization between them, leading to a small mass (gap) term. The
arguments based on Figs.~\ref{fig5}(a)-\ref{fig5}(c) give a reason for
the smallness of the induced mass gap. 
If the spin-orbit coupling is neglected, the two bands
forming the Dirac electrons belong to different irreducible
representations on the $\Gamma$-X line, so that a gap does not
open. Namely, the spin-orbit
coupling is necessary to make the mass term finite, even if
$d_{zx/yz}$ or $d_{xy}$ orbitals are included in the tight-binding model.

We discuss some points hereafter. The first is about related
materials. We studied the band structures of the family of Ca$_3$PbO,
namely, Sr$_3$PbO, Ba$_3$PbO, and Ca$_3$SnO. (Figures are not shown
here.) In the calculation, the lattice constants of these materials are
again taken from experimental results\cite{Widera19801805}. 
We find that all these systems have similar band structures and Dirac
electrons. We point out here that the overlap of the $p$- and
$d$-bands, which is essential for the existence of Dirac electrons,
becomes larger in the order Ca$_3$PbO $\rightarrow$ Sr$_3$PbO
$\rightarrow$ Ba$_3$PbO. However, in
Ba$_3$PbO, some other (non-Dirac) bands appear at the Fermi energy,
which may mask the properties of the Dirac electrons.
For Ca$_3$SnO, the overlap of the $p$- and $d$-bands becomes smaller
than that in Ca$_3$PbO, but it still exists and gives Dirac
electrons at the Fermi energy. 
Since the spin-orbit coupling is smaller for Sn than for Pb, the
estimated mass gap is actually as small as $4$ meV, compared with $14$
meV for Ca$_3$PbO. From this result, we propose that the series of alloys
Ca$_3$(Pb$_{1-x}$Sn$_x$)O will provide a method of controlling the mass
term of Dirac electrons. 

Compared with Bi, which has three-dimensional
massive Dirac electrons in its low energy band structure, the present
system has a simple structure whose Fermi energy is located in the
gap of the Dirac electrons. In Bi, on the other hand, there is a
complication due to its
semimetallic properties\cite{interband}. As a result, fine-tuning of
the doping or
pressure is necessary for studying the Dirac electrons in Bi. Ca$_3$PbO and
its family will be interesting and important materials for exploring the
three-dimensional Dirac electrons.

Responses to applied magnetic fields will also be
interesting. Neglecting the small gap, a Dirac electron has a linear
dispersion, leading to a peculiar Landau level structure\cite{Zhang:2005kx}. 
Another possible interesting response is a large orbital diamagnetism,
which is observed in Bi\cite{wehrli,JPSJ.28.570}. The
orbital diamagnetism takes a maximum when the Fermi energy is {\it
inside} the gap, namely, a large electronic response is observed even
though there are no Fermi surfaces. The interesting interband effects in
the quantum transport phenomenon are being extensively
studied\cite{Fuseya}.

If a small number of electrons (or holes) are doped to this material, it
becomes a metal with six tiny Fermi surfaces, or six {\it
valleys}. It will be interesting to consider the possibility of
a three-dimensional analogue of {\it valleytronics}, which has recently been
developed for graphene\cite{Rycerz:2007vn,PhysRevLett.106.136806}. It is also
interesting to consider the
possibility of spontaneous valley symmetry breaking, in which the
occupancy of the valleys spontaneously becomes imbalanced. 

In summary, we have performed the band structure calculation of
Ca$_3$PbO and its family, and found that three-dimensional
massive Dirac electrons exist at the Fermi energy. A tight-binding model that
describes the essence of the emergence of Dirac electrons was
successfully constructed. By analyzing this model, it is concluded that
the symmetry of the crystal and the involved orbitals play important
roles in sustaining Dirac electrons in the band structure.

\begin{acknowledgments}
The authors appreciate H.~Fukuyama and Y.~Fuseya for stimulating discussions.
T.K. was supported by JSPS Research Fellowship. 
\end{acknowledgments}


\begin{thebibliography}{10}
\bibitem{comment1} Codes to analyze the orbital weight and the
	irreducible representation of the bands are included in the WIEN2k
	package.
\end{thebibliography}
\end{document}